\documentclass[twocolumn,showpacs,amsfonts,aps,prc,floatfix,%
superscriptaddress]{revtex4} 
\usepackage{epsfig}
\usepackage{bm}

\newcommand{\nc}{\newcommand}

\nc{\2}{\frac{1}{2}}
\nc{\half}{{\textstyle\2}}
\nc{\3}{\frac{1}{3}}
\nc{\4}{\frac{1}{4}}

\nc\be{\begin{equation}}
\nc\ee{\end{equation}}
\nc{\beq}[1]{\begin{equation}\label{#1}}
\nc{\eeq}{\end{equation}}
\nc{\bea}[1]{\begin{eqnarray}\label{#1}}
\nc{\eea}{\end{eqnarray}}

\nc{\bce}{\begin{center}}
\nc{\ece}{\end{center}}
\nc{\bit}{\begin{itemize}}
\nc{\eit}{\end{itemize}}
\nc{\bmp}{\begin{minipage}}
\nc{\emp}{\end{minipage}}
\nc{\se}{\section}
\nc{\suse}{\subsection}

\nc{\bb}{\bm{b}}
\nc{\bq}{\bm{q}}
\nc{\bp}{\bm{p}}
\nc{\bK}{\bm{K}}
\nc{\br}{\bm{r}}
\nc{\bs}{\bm{s}}

\nc{\Eq}{{\,=\,}}
\nc{\Kt}{K_\perp}
\nc{\pt}{p_\perp}
\nc{\mt}{m_\perp}
\nc{\pL}{p_{\rm L}}
\nc{\ET}{E_{\rm T}}
\nc{\Nch}{N_{\rm ch}}
\nc{\Nc}{N_{\rm coll}}
\nc{\Np}{N_{\rm part}}
\nc{\Atanh}{{\rm Atanh}}
\nc{\Asinh}{{\rm Asinh}}
\nc{\Acosh}{{\rm Acosh}}
\nc{\scm}{\sqrt{s_{\rm NN}}}

\newcommand{\gapp}{\raisebox{-.5ex}{$\stackrel{>}{\sim}$}}
\newcommand{\lapp}{\raisebox{-.5ex}{$\stackrel{<}{\sim}$}}

\nc{\la}{\langle}
\nc{\lla}{\left \langle}
\nc{\ra}{\rangle}
\nc{\rra}{\right \rangle}

\begin{document}

\title{Projected Three-Pion Correlation Functions}

\author{Ulrich Heinz}
\email[E-mail: ]{heinz@mps.ohio-state.edu}
\affiliation{Physics Department, The Ohio State University, Columbus, OH 43210}
\author{Alex Sugarbaker}
\email[E-mail: ]{alexs@uchicago.edu}
\affiliation{Physics Department, The University of Chicago, 
             Chicago, IL 60637}

%
%

\begin{abstract}
We propose a new procedure for constructing projected three-pion
correlation functions which reduces undesirable artificial
momentum dependences resulting from the commonly used procedure 
and facilitates comparison of three-pion correlation data with 
theoretical models.
\end{abstract}

\vspace{1.0cm}

\date{\today}

\maketitle

%
%

\section{Introduction}
\label{sec1}

Two-particle intensity interferometry, exploiting Bose-Einstein 
correlations between pairs of identical bosons (pions, kaons), has 
been extensively used to extract information about the space-time 
structure of high-energy hadron-hadron and nucleus-nucleus collisions 
\cite{WH99}. Three-particle Bose-Einstein correlations were shown to 
yield additional information which can not be extracted from 
two-particle correlations \cite{APW93,HV96,HZ97,HV97}: (i) After 
subtracting the two-particle correlation contributions and 
extrapolating to zero relative momentum, the strength of the 
genuine three-particle correlation term relative to the two-particle 
correlator provides an unambiguous measure for the degree of 
chaoticity of the source (or, conversely, it limits the degree 
of phase-coherence in the particle emission process) \cite{HZ97}. 
This idea has recently been applied by the STAR Collaboration 
\cite{STAR3pi} to three-pion correlations from 130\,$A$\,GeV Au+Au 
collisions at RHIC, showing that at these collision energies the 
data are consistent with a completely incoherent (``chaotic'') pion 
emission process. (ii) For a completely chaotic source, the 
full momentum dependence of the ``reduced'' three-pion correlator 
$r_3(\bp_1,\bp_2,\bp_3)$ can be used to extract spatial asymmetries 
of the source around its point of maximum emissivity \cite{HZ97,HV97}; 
for a partially coherent source, the momentum dependence of $r_3$ 
constrains the relative spatial sizes of the coherent and incoherent 
parts of the emission function \cite{HZ97}. This information is, 
however, difficult to extract because the required accurate analysis 
of its full momentum dependence puts statistical demands on the
measured three-pion correlator which can not be met by presently available
data.

Statistical limitations therefore so far force experimentalists to 
project the three-pion correlation function onto a single relative
momentum variable. The preferred choice \cite{HV96,STAR3pi,NA44_3pi,WA98_3pi} 
for this variable is a Lorentz scalar that is completely symmetric under 
interchange of the three pions: $Q_3\Eq\sqrt{Q^2_{12}{+}Q^2_{23}{+}Q^2_{31}}$
where $Q^2_{ij}\Eq{-}(p_i{-}p_j)^2$ are the Lorentz-invariant relative
momenta between pairs in the pion triplet. This projection has
undesirable, but to some extent unavoidable consequences for the 
relative momentum dependence of the three-pion correlator. 
Specifically, it is known \cite{HZ97} that for a completely chaotic 
source the leading relative momentum dependence of 
$r_3(\bK,\bq_{12},\bq_{23})$ (where $\bK\Eq(\bp_1{+}\bp_2{+}\bp_3)/3$
is the average momentum of the triplet while $\bq_{12}\Eq\bp_1{-}\bp_2$
and $\bq_{23}\Eq\bp_2{-}\bp_3$ are two linearly independent relative
momenta) at small $q_{ij}$ is of $4^{\rm th}$ order in the components 
of the relative momenta $\bq_{ij}$. The projection on $Q_3$ instead 
introduces a dominant quadratic $q$-dependence which buries the 
coefficients of the $4^{\rm th}$- and higher-order contributions that 
would allow to extract additional interesting source information, such 
as the momentum dependence of the point of maximum emissivity and the 
asymmetry of the source around that point \cite{HZ97}.

In this short note we investigate in some detail the effects of this 
projection procedure. We show that there are different possibilities
to project onto a single variable $Q_3$ some of which exacerbate the
above problem far beyond the unavoidable minimum. In particular, the
version employed so far in the data analysis has the undesirable 
feature of introducing a relative momentum dependence in the projected 
three-pion correlator {\em even for a static Gaussian source for which the 
underlying unprojected correlator is completely momentum-independent}. 
We therefore suggest a different projection method which avoids this 
undesirable feature and furthermore allows for a ``staged projection'' 
onto any subset or different combination of the three particle momenta. 
As a side effect, the new method also facilitates the comparison of 
the projected three-pion correlator with theoretical models.

\section{Two- and three-particle correlators}
\label{sec2}

The two- and three-particle correlation functions are defined as the 
ratios between the two- and three-particle coincidence cross sections 
and the products of independent single-particle cross sections:
\bea{1}
  &&C_2(12) \equiv C_2(\bp_1,\bp_2) 
  = \frac{N_2(\bp_1,\bp_2)}{N_1(\bp_1)N_1(\bp_2)}\,,
\\\nonumber
  &&C_3(123) \equiv C_3(\bp_1,\bp_2,\bp_3) 
  = \frac{N_3(\bp_1,\bp_2,\bp_3)}{N_1(\bp_1)N_1(\bp_2)N_1(\bp_3)}\,,
\eea
where
\bea{2}
  N_1(\bp_i) &=& \frac{d^3N}{d^3p_i}\quad (i=1,2,3)\,,
\nonumber\\
  N_2(\bp_1,\bp_2) &=& \frac{d^6N}{d^3p_1 d^3p_2}\,,
\nonumber\\
  N_3(\bp_1,\bp_2,\bp_3) &=& \frac{d^9N}{d^3p_1 d^3p_2 d^3p_3}\,.
\eea
Equations (\ref{1}) are frame-independent even though they don't look
that way since the energy factors $E_i$ making the cross sections
$E_i dN/d^3p_i$ etc. Lorentz invariant cancel between numerators and 
denominators. In the absence of correlations $C_2\Eq{C_3}\Eq1$. One 
defines the ``true $n$-particle correlators'' $R_n$ (cumulants) by 
subtracting uncorrelated parts and correlations involving fewer 
than $n$ particles: $R_2(ij){\Eq}C_2(ij){-}1$ and 
$R_3(123){\Eq}C_3(123)-R_2(12){-}R_2(23){-}R_2(31){-}1$.
The ``normalized true three-particle correlator'' 
$r_3(123)\Eq{r_3}(\bp_1,\bp_2,\bp_3)$ \cite{HZ97} is obtained by 
dividing $R_3(123)$ by the square root of the product of 2-particle 
correlators:
\begin{widetext}
\bea{3}
 r_3(123) = \frac{\bigl(C_3(123){-}1\bigr) - \bigl(C_2(12){-}1\bigr)
                    - \bigl(C_2(23){-}1\bigr)- \bigl(C_2(31){-}1\bigr)}
                    {\sqrt{\bigl(C_2(12){-}1\bigr)\bigl(C_2(23){-}1\bigr)
                           \bigl(C_2(31){-}1\bigr)}}.
\eea
\end{widetext}
For a fully chaotic source, $r_3$ approaches the value 2 at zero 
relative momenta, $\bq_{12}\Eq\bq_{23}\Eq0$, for any value of the
triplet momentum $\bK\Eq(\bp_1{+}\bp_2{+}\bp_3)/3$. The leading deviations
from this limit in both numerator and denominator are of second order 
in the components of $\bq_{ij}$, but these leading corrections cancel
in the ratio, leaving a leading $4^{\rm th}$ order term in $r_3$ 
\cite{HZ97}. It is obvious that a projection of the numerator and 
denominator onto a smaller number of momentum components will in 
general destroy the cancellation of the leading $2^{\rm nd}$ order
terms, thus generating an artificial leading quadratic 
$\bq_{ij}$-dependence of the projected $r_3$. This leading quadratic 
$q$-dependence is not interesting because its coefficients depend on 
properties of the source which can already be extracted from the 
two-particle correlation function. Unfortunately, this extraction is
not completely model-independent \cite{WH99}, so it is not clear
whether, even in principle, it might be possible to subtract the
unwanted second order terms, using the measured two-particle correlators, 
in order to uncover the subleading $4^{\rm th}$ order terms in the 
projected $r_3$. We will not address this issue here. Instead we will 
focus on a different question: {\em Is it possible to minimize these 
projection artefacts by using an optimized projection algorithm?}

\vspace*{-1cm}

\section{Shortcomings of the ``standard'' projection method}
\label{sec3}

In \cite{STAR3pi} the STAR collaboration constructed a projected 
version $\bar r_3(Q_3)$ of the normalized true three-pion correlator
using the following procedure: First, they constructed projected
two- and three-particle correlation functions, $\bar C_2(Q_{ij})$
and $\bar C_3(Q_3)$, by taking pairs and triplets of pions from 
the same collision event, summed over all events, binning them
in the Lorentz-invariant variables $Q_{ij}$ and $Q_3$, respectively,
and dividing the contents of each bin by that of an equivalent bin
at the same $Q$-value which was filled with uncorrelated pairs 
or triplets generated from mixed events \cite{STAR3pi}. This 
procedure can be formally represented by the following definitions:
\begin{widetext}
\bea{4}
 \bar C_2(Q_{ij}) &=& 
 \frac{\int d^3p_i d^3p_j \,N_2(\bp_i,\bp_j)\,
            \delta\left(Q_{ij}-\sqrt{-(p_i{-}p_j)^2}\right)}
      {\int d^3p_i d^3p_j \,N_1(\bp_i) N_1(\bp_j)\,
            \delta\left(Q_{ij}-\sqrt{-(p_i{-}p_j)^2}\right)},
 \qquad (i,j=1,2,3)
\\
\label{5}
 \bar C_3(Q_3) &=& 
 \frac{\int d^3p_1 d^3p_2 d^3p_3\, N_3(\bp_1,\bp_2,\bp_3)\,
 \delta\left(Q_3-\sqrt{-(p_1{-}p_2)^2-(p_2{-}p_3)^2-(p_3{-}p_1)^2}\right)}
      {\int d^3p_1 d^3p_2 d^3p_3\, N_1(\bp_1) N_1(\bp_2) N_1(\bp_3)\,
 \delta\left(Q_3-\sqrt{-(p_1{-}p_2)^2-(p_2{-}p_3)^2-(p_3{-}p_1)^2}\right)}.
\eea
From these projected correlation functions one then constructs the
ratio (\ref{3}) by fixing $Q_3$ and summing over all two-particle
relative momenta $Q_{ij}$ whose squares add up to $Q_3^2$. Formally
\bea{6}
 &&\bar r_3(Q_3) = \left[\int_0^\infty dQ_{12} dQ_{23} dQ_{31} \, 
   \delta\Bigl(Q_3-\sqrt{Q_{12}^2{+}Q_{23}^2{+}Q_{31}^2}\Bigr)\right]^{-1}
   \times
\nonumber\\
 &&\int_0^\infty dQ_{12} dQ_{23} dQ_{31} \,
              \delta\Bigl(Q_3-\sqrt{Q_{12}^2{+}Q_{23}^2{+}Q_{31}^2}\Bigr)
                 \frac{\bigl(\bar C_3(Q_3){-}1\bigr) 
                     - \bigl(\bar C_2(Q_{12}){-}1\bigr)
                     - \bigl(\bar C_2(Q_{23}){-}1\bigr)
                     - \bigl(\bar C_2(Q_{31}){-}1\bigr)}
                      {\sqrt{\bigl(\bar C_2(Q_{12}){-}1\bigr)
                             \bigl(\bar C_2(Q_{23}){-}1\bigr)
                             \bigl(\bar C_2(Q_{31}){-}1\bigr)}}.
\quad
\eea
\end{widetext}
Note that the four terms in the numerator of the integrand in 
Eq.~(\ref{6}) involve four different $\delta$-functions. This is 
clearly inconvenient when comparing to theoretical model calculations 
(for example those by Nakamura and Seki \cite{NS99}) which are not 
easily projected in the same way. More seriously, however, the 
mixing of different projection algorithms for $\bar C_3$ and 
$\bar C_2$ in the numerator of Eq.~(\ref{6}) upsets the above-mentioned
partial cancellation of the relative momentum dependences in the numerator
and denominator of the integrand. It thereby acerbates the problem
of introducing unwanted $q$-dependence into $r_3$ by projecting it.
 
We demonstrate this by considering a very simple, spherically symmetric 
source characterized by the following (unnormalized) emission function:
\bea{7}
   S(x,p) = E_p\, \delta(t{-}t_0)\,e^{-\br^2/2R^2}\, e^{-\bp^2/2\Delta^2}.
\eea
Particles are emitted instantaneously at global time $t_0$,
with a Gaussian momentum distribution which is independent of the 
emission point (no ``$x$-$p$-correlations''). The factor 
$E_p\Eq\sqrt{m^2{+}\bp^2}$ is introduced for calculational 
convenience so that it drops out from relations such as 
$E_p\,dN/d^3p\Eq\int d^4x\,S(x,p)$ which relate the $n$-particle 
spectra $N_n(\bp_1,\dots,\bp_n)$ in Eq.~(\ref{2}) to the emission 
function $S(x,p)$ \cite{WH99,APW93,HV96,HZ97}. For simplicity we 
assume nonrelativistic momenta, $\Delta{\,<\,}m$. General arguments
presented in \cite{HZ97} suggest that for such a source the exact
reduced three-particle correlator $r_3$ from Eq.~(\ref{3})
should be equal to 2 and completely momentum-independent. The 
(nonrelativistic) two-particle exchange amplitude 
$\rho_{ij}\Eq\langle\hat a^\dagger(\bp_i)\hat a(\bp_j)\rangle$
\cite{APW93,CH94,fn1} for the source (\ref{7}) reads
\bea{8}
   \rho_{ij} &=& \int d^4x\, \frac{S(x,K_{ij})}{\sqrt{E_{K_{ij}}}}
   \, e^{iq_{ij}\cdot x}
\nonumber\\
   &=& (2\pi R^2)^{\frac{3}{2}} e^{-\frac{\bK_{ij}^2}{2\Delta^2}
                         -\frac{\bq_{ij}^2 R^2}{2}}
   e^{i(E_i-E_j)t_0},
\eea
where $\bK_{ij}\Eq(\bp_i{+}\bp_j)/2$ and $\bq_{ij}\Eq\bp_i{-}\bp_j$
are the average and relative momentum of the pair $ij$. It allows 
to calculate the one-, two- and three-particle cross sections from 
the relations \cite{APW93,HZ97}
\bea{9}
   &&N_1(i) = \rho_{ii},\quad
     N_2(ij) = \rho_{ii}\rho_{jj} + |\rho_{ij}|^2,
\nonumber\\
   &&N_3(123) = 2\, {\rm Re} (\rho_{12}\rho_{23}\rho_{31}) + 
   |\rho_{12}|^2\rho_{33}+|\rho_{23}|^2\rho_{11}
\nonumber\\
   &&\qquad\qquad\ \ 
      + |\rho_{31}|^2\rho_{22}+\rho_{11}\rho_{22}\rho_{33}
\eea
(where $N_1(i)$ is shorthand for $N_1(\bp_i)$, etc.), yielding
\bea{10} 
  N_1(i)N_1(j)&=&(2\pi R^2)^3\, e^{-\frac{\bK_{ij}^2}{\Delta^2}}
  \, e^{-\frac{\bq_{ij}^2}{4\Delta^2}},
\\\nonumber
  N_2(ij) &=& (2\pi R^2)^3\, e^{-\frac{\bK_{ij}^2}{\Delta^2}}
  \left(e^{-\frac{\bq_{ij}^2}{4\Delta^2}} + e^{-\bq_{ij}^2 R^2}\right),
\eea
and
\begin{widetext}
\bea{11}
  N_1(1)N_1(2)N_1(3) &=& (2\pi R^2)^{\frac{9}{2}}\, 
  e^{-\frac{3}{2}\frac{\bK^2}{\Delta^2}}\,
  e^{-\frac{\bq_{12}^2+\bq_{23}^2+\bq_{12}\cdot\bq_{23}}{3\Delta^2}},
\\\label{12}
  N_3(123) &=& (2\pi R^2)^{\frac{9}{2}}\,
  e^{-\frac{3}{2}\frac{\bK^2}{\Delta^2}}\,
  \left(2\,e^{-(\bq_{12}^2+\bq_{23}^2+\bq_{12}\cdot\bq_{23})
               \left(R^2+\frac{1}{12\Delta^2}\right)} 
  + e^{-\frac{(\bq_{23}{-}\bq_{31})^2}{12\Delta^2}-R^2(\bq_{23}{+}\bq_{31})^2}
  \right.
\\\nonumber
  &&\qquad\qquad\left.
  + e^{-\frac{(\bq_{31}{-}\bq_{12})^2}{12\Delta^2}-R^2(\bq_{31}{+}\bq_{12})^2}
  + e^{-\frac{(\bq_{12}{-}\bq_{23})^2}{12\Delta^2}-R^2(\bq_{12}{+}\bq_{23})^2}
  + e^{-\frac{\bq_{12}^2+\bq_{23}^2+\bq_{12}\cdot\bq_{23}}{3\Delta^2}}
   \right).
\eea
\end{widetext}  
After plugging these into equations (\ref{1}) and (\ref{3}), a bit of 
algebra confirms that the exact $r_3$ is indeed momentum-independent 
and equal to $r_3\Eq2$.

\begin{figure}
\includegraphics[width=0.95\linewidth]{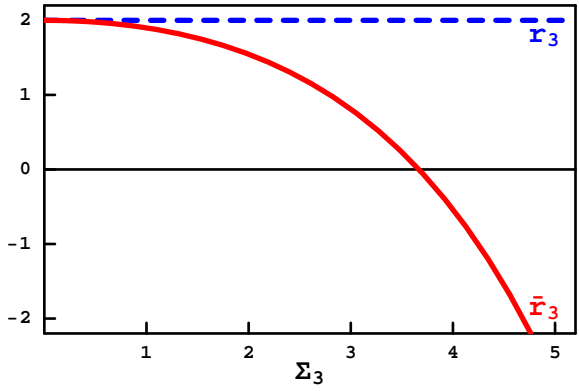}
\caption{\label{F1}(color online) The normalized true three-particle 
correlator $r_3$ (dashed, blue) and its projected version $\bar r_3$ 
(solid, red) from Eq.~(\ref{6}) for the simple source (\ref{7}), as 
functions of the dimensionless variable 
$\Sigma_3{\Eq}Q_3^2\left(R^2{-}1/(4\Delta^2)\right)$.} 
\end{figure}

Let us now study the projected correlator $\bar r_3(Q_3)$ defined in
Eq.~(\ref{6}). For non-relativistic momenta we can use 
$-(p_i{-}p_j)^2{\,\approx\,}(\bp_i{-}\bp_j)^2\Eq\bq_{ij}^2$ which allows
to do the integrals in Eqs.~(\ref{4},\ref{5}) analytically. 
After transforming integration variables according to 
$d^3p_i d^3p_j{\Eq}d^3K_{ij} d^3q_{ij}$
and $d^3p_1 d^3p_2 d^3p_3{\Eq}d^3K d^3q_{ij} d^3q_{jk}$ (where
$(ij),(jk)$ indicate two of the three possible pairs (12),\,(23),\,(31), 
as suitable to exploit the cyclic symmetry of the three middle terms in 
Eq.~(\ref{12})) and further $d^3q_{ij} d^3q_{jk}{\Eq}d^3\xi d^3\zeta$ 
where $\bm{\zeta}\Eq\bq_{ij}{-}\bq_{jk}$ and 
$\bm{\xi}\Eq(\bq_{ij}{+}\bq_{jk})/2$, the integrations are straightforward, 
and we obtain for the projected correlators (\ref{4},\ref{5})
\bea{13}
  \bar C_2(Q_{ij})-1 &=& e^{-\Sigma_{ij}},
\\
\label{14}
  \bar C_3(Q_3)-1 &=& 2\, e^{-\frac{1}{2}\Sigma_3} + \frac{48}{\pi}
  \int_0^1 dz\,z^2 \sqrt{1{-}z^2}\,e^{-\frac{2}{3}\Sigma_3 z^2}
\nonumber\\
  &=& 2\, e^{-\frac{1}{2}\Sigma_3} + 6 \, e^{-\frac{1}{3}\Sigma_3}
  \, \frac{{\rm I}_1\left(\frac{1}{3}\Sigma_3\right)}{\frac{1}{3}\Sigma_3}.
\eea      
Here $z\Eq\cos\phi$ arises from the angle between $\bm{\xi}$
and $\bm{\zeta}$, I$_1$ is a modified Bessel function, and
\bea{15}
  \Sigma_{ij} &=& Q_{ij}^2\left(R^2-\frac{1}{4\Delta^2}\right),
\nonumber\\
  \Sigma_3 &=& \Sigma_{12}+\Sigma_{23}+\Sigma_{31} 
            = Q_3^2\left(R^2-\frac{1}{4\Delta^2}\right).
\eea
The emission function $S(x,p)$ is a Wigner density \cite{WH99} whose 
spatial and momentum widths must satisfy the uncertainty relation
$R^2 \Delta^2 \geq (\hbar/2)^2$. Hence, $\Sigma_{ij}$ and $\Sigma_3$
are positive quantities. Inserting Eqs.~(\ref{13},\ref{14}) back into 
Eq.~(\ref{6}) and introducing spherical polar coordinates, one angular 
integral can be done trivially while the radial integral is executed 
with the help of the $\delta$-function. The remaining integral over 
the polar angle yields an error function:
\bea{16}
\nonumber
  \bar r_3(Q_3) &=& 2 + 3 \biggl[ e^{\frac{1}{2}(\Sigma_3/3)}
  \frac{{\rm I}_1\left(\Sigma_3/3\right)}
  {\frac{1}{2}\left(\Sigma_3/3\right)}
\\
  &&\quad\quad
  - \frac{\sqrt{\pi}}{2} e^{\frac{1}{2}\Sigma_3} 
    \frac{{\rm erf}\left(\sqrt{\Sigma_3}\right)}{\sqrt{\Sigma_3}}
  \biggr].
\eea
$\bar r_3$ is plotted in Figure~\ref{F1} as a function of $\Sigma_3$
in the relevant range $\Sigma_3{\,\lapp\,}3$ where the two- and 
three-particle correlation functions $\bar C_2(Q_{ij})$ and 
$\bar C_3(Q_3)$ show measurable deviations from unity 
\cite{STAR3pi,NA44_3pi,WA98_3pi}.
One observes a significant discrepancy to the exact result $r_3\Eq2$, 
and for large $\Sigma_3$ one checks that $\bar r_3$ grows 
exponentially negative: 
$\bar r_3{\,\to\,}{-}\frac{3}{2}\sqrt{\pi e^{\Sigma_3}/\Sigma_3}$.
At small relative momenta its leading $Q_3$-dependence is
\bea{17}
  \bar r_3(Q_3) = 2 - \frac{11}{120} \Sigma_3^2 + {\cal O}(\Sigma_3^3),
\eea
i.e. of 4$^{\rm th}$ order in the $\bq_{ij}$. The continued 
cancellation of 2$^{\rm nd}$ order terms is probably accidental and
due to the particularly simple structure of the emission function
(\ref{7}). The coefficient in front of the $\Sigma_3^2$ term is small,
and the deviations of $\bar r_3$ from 2 become significant only
at $\Sigma_3{\,\gapp\,}1$. This may explain why Humanic did not
see such a deviation in his Monte Carlo model study 
\cite{Humanic:1999ns} which explored $\bar r_3$ in a restricted $Q_3$ 
range and with limited statistical accuracy, even though he employed 
the same projection procedure (\ref{6}). 

\section{A different projection method}
\label{sec4}

The $Q_3$-dependence in Eq.~(\ref{16}) is entirely artificial and 
generated by using different projection procedures in equations 
(\ref{4}) and (\ref{5}) and mixing them in Eq.~(\ref{6}). It can 
be easily avoided, however. Inserting the definitions (\ref{1}) into 
Eq.~(\ref{3}) and multiplying both numerator and denominator by 
$N_1(\bp_1)N_1(\bp_2)N_1(\bp_3)$, $r_3$ can be brought into the form
\begin{widetext}
\bea{18}
  r_3(123) &=& 
  \frac{N_3(\bp_1,\bp_2,\bp_3) - N_2(\bp_1,\bp_2) N_1(\bp_3)
      - N_2(\bp_2,\bp_3) N_1(\bp_1) - N_2(\bp_3,\bp_1) N_1(\bp_2)
      + 2 N_1(\bp_1)N_1(\bp_2)N_1(\bp_3)}
       {\sqrt{[N_2(\bp_1,\bp_2)-N_1(\bp_1)N_1(\bp_2)]
              [N_2(\bp_2,\bp_3)-N_1(\bp_2)N_1(\bp_3)]
              [N_2(\bp_3,\bp_1)-N_1(\bp_3)N_1(\bp_1)]}}
\nonumber\\
  &\equiv& \frac{[{\rm NUM}(\bp_1,\bp_2,\bp_3)]}
           {[{\rm DEN}(\bp_1,\bp_2,\bp_3)]}.
\eea
The numerator contains sums and differences of different kinds of
triplet yields: the first term denotes real triplets where all three
pions come from the same collision event, the last term mixed event 
triplets where all three pions come from different events, and the 
middle three terms subtract mixed-event triplets where two pions come 
from the same and the third from a different collision. For consistency
with Eqs.~(\ref{9}), the distributions of both types of mixed-event 
triplets must be normalized such that they agree with the real triplets 
at large relative momenta where all quantum statistical correlations
disappear: Eq.~(\ref{8}) shows that for $i{\,\ne\,}j$, $\rho_{ij}$ 
vanishes in the limit $|\bq_{ij}|{\,\to\,}\infty$. 

The denominator in (\ref{18}) involves a product of differences
between yields of real pairs and mixed-event pairs (again normalized
to the same total number of pairs), with pion momenta as indicated
by the arguments. It can be rewritten as the yield of mixed-event 
triplets multiplied by a weight factor,
\bea{19}
  [{\rm DEN}(\bp_1,\bp_2,\bp_3)] = N_1(\bp_1)N_1(\bp_2)N_1(\bp_3)
  \cdot {\cal W}(\bp_1,\bp_2,\bp_3),
\eea
where the latter is computed from the (Coulomb corrected) two-particle 
correlation function $C_2(\bp_i,\bp_j)$ according to
\bea{20}
  {\cal W}(\bp_1,\bp_2,\bp_3) = 
  \sqrt{[C_2(\bp_1,\bp_2)-1][C_2(\bp_2,\bp_3)-1][C_2(\bp_3,\bp_1)-1]}.
\eea
Fully six-dimensional analyses of the two-particle correlation
function $C_2(\bp_i,\bp_j){\Eq}C_2(\bK_{ij},\bq_{ij})$ are available.
It can be tabulated and interpolated, allowing for a straightforward
calculation of the weight ${\cal W}(\bp_1,\bp_2,\bp_3)$. (Again the
mixed-pair background must be normalized such that 
$C_2(|\bq_{ij}|{\to}\infty)\Eq1$.)

We propose to construct $r_3$ directly from the ratio (\ref{18}) 
instead of first constructing two- and three-particle correlation 
functions and then using (\ref{3}). It is straightforward to project 
Eq.~(\ref{18}) onto the single relative momentum variable $Q_3$ (or 
onto any other desired combination of the three pion momenta) by
binning the numerator and denominator separately in $Q_3$ and
taking the ratio:
\bea{21}   
  \langle r_3\rangle(Q_3) =
  \frac{\int d^3p_1 d^3p_2 d^3p_3\,\bigl[{\rm NUM}(\bp_1,\bp_2,\bp_3)\bigr]\,
  \delta\left(Q_3-\sqrt{-(p_1{-}p_2)^2-(p_2{-}p_3)^2-(p_3{-}p_1)^2}\right)}
       {\int d^3p_1 d^3p_2 d^3p_3\,\bigl[{\rm DEN}(\bp_1,\bp_2,\bp_3)\bigr]\,
  \delta\left(Q_3-\sqrt{-(p_1{-}p_2)^2-(p_2{-}p_3)^2-(p_3{-}p_1)^2}\right)}.
\eea
\end{widetext}
Note that, by symmetry, the three middle terms in the numerator of 
Eq.~(\ref{18}) give identical contributions to Eq.~(\ref{21}). The 
numerator is thus obtained by first running separately through all 
real triplets, all triplets with two particles from the same event 
and the third particle from a different event, and all mixed-event
triplets, binning each of these in $Q_3$. After normalizing the bin
contents of the mixed-event triplets by the appropriate factor which 
ensures that their distributions agree with that of the real triplets 
at large $|\bq_{ij}|$ (note that this does {\em not} imply that they
necessarily agree at large $Q_3$!), one then subtracts, for each value 
of $Q_3$, from the bin containing the real triplets three times the 
bin content from the half-mixed triplets and adds twice the content 
from the fully mixed triplets. For the denominator one runs over all 
fully mixed triplets, calculates for each of them the weight ${\cal W}$ 
and bins these weights in $Q_3$. At the end one normalizes the bin 
contents by the same factor as for the fully mixed triplets in the 
numerator.

While this projection procedure does not, of course, completely sidestep 
the unavoidable interference with the cancellation of the leading relative 
momentum dependences between numerator and denominator, at least it
preserves the {\em momentum-independence} of $r_3$ in the ``trivial'' 
case (\ref{7}) of a spherically symmetric, static source without 
$x$-$p$-correlations: For the source (\ref{7}), 
$\langle r_3\rangle(\bp_1,\bp_2,\bp_3){\,\equiv\,}r_3(\bp_1,\bp_2,\bp_3)%
{\,\equiv\,}2$. 

A technical complication for the data analysis arises from the repulsive 
final state Coulomb interaction between the emitted charged pions which 
must be corrected for before computing the reduced three-pion correlator
$r_3$. This problem poses itself similarly for both projection 
techniques and can be dealt with in the same way in either approach,
by using Coulomb correction factors extracted from a self-consistent
Coulomb correction of the two-pion correlation functions. We refer to 
Refs.~\cite{WA98_3pi,STAR3pi} for details. Note that in the form given 
in Eq.~(\ref{18}), the Coulomb correction must be applied to the
{\em real} \ pairs and triplets, and not to the mixed-event ones. 

\medskip

\section{Conclusions}
\label{sec5}

We have shown that, for a simple spherically symmetric, non-expanding
source with sudden particle emission, the standard projection procedure,
as up to now applied in the experimental analysis of three-pion 
correlations, introduces an artificial momentum dependence into
the normalized true three-pion correlator $r_3$, even though 
the exact result for $r_3$ is completely momentum-independent. This 
suggests that the observed $Q_3$ dependence of $\bar r_3$ seen by 
WA98 in Pb+Pb collisions at the SPS \cite{WA98_3pi} and by STAR in 
Au+Au collisions at RHIC \cite{STAR3pi} may be significantly affected 
by artificial projection effects. Even though projection-induced 
relative momentum dependence of $r_3$ can in general not be 
completely avoided, it can be minimized by using the new projection 
procedure suggested in this paper. This procedure computes $r_3$ 
in a more direct way and applies the same binning to all terms. It 
preserves the momentum-independence of $r_3$ for a symmetric 
non-expanding source. The new projection algorithm can also be easily 
employed in theoretical model calculations, rendering the comparison 
with data (but not necessarily the interpretation of the extracted 
source parameters) more straightforward.

\vspace*{-5mm}

\begin{acknowledgments}
UH gratefully acknowledges clarifying discussions with Robert Willson 
and Tom Humanic about the projection method used by the NA44 and STAR 
Collaborations. We thank Mikhail Kopytine for pointing out a normalization 
error in the first version of the manuscript. AS thanks the Physics 
Department at Ohio State University for hospitality during their 2004 
REU program. This work was supported by the U.S. Department of Energy 
under contract DE-FG02-01ER41190 and by NSF through REU grant PHY-0242665.
\end{acknowledgments}



\begin{thebibliography}{99}
\bibitem{fn1}
  Note that we here use for simplicity a non-covariant, nonrelativistic 
  form of the two-particle exchange amplitude which differs from the
  covariant one \cite{APW93,CH94} by the factor 
  $\sqrt{E_iE_j}{\,\approx\,}E_{K_{ij}}$.
\end{thebibliography}
\end{document}